\documentclass[]{elsarticle}
\usepackage[english]{babel}
\usepackage{graphicx}
\usepackage{lineno}
\usepackage{amssymb,amsmath}

\usepackage[]{hyperref} 	
\usepackage[]{epsfig}
\usepackage{dcolumn}
\usepackage{epstopdf}
\usepackage{multirow}

\newcommand{\queens}{Department of Physics, Engineering Physics \& Astronomy, Queen's University, Kingston, ON, Canada, K7L 3N6}

\newcommand{\kvi}{Kernfysisch Versneller Instituut, University of Groningen, NL-9747AA Groningen, The Netherlands}

\newcommand{\Mb}{\mbox{$\mbox{Mbytes}$}}
\newcommand{\Gb}{\mbox{$\mbox{Gbytes}$}}
\newcommand{\kelv}{\mbox{$\mbox{K}$}}
\newcommand{\mK}{\mbox{$\mbox{mK}$}}
\newcommand{\zwo}{\mbox{$\mbox{ZnWO}_4$}}

\newcommand{\cs}{\mbox{${}^{137}\mbox{Cs}$}}
\newcommand{\am}{\mbox{${}^{241}\mbox{Am}$}}
\newcommand{\ttna}{\mbox{${}^{22}\mbox{Na}$}}
\newcommand{\keV}{\mbox{$\mbox{keV}$}}
\newcommand{\micros}{\mbox{$\mu\mbox{s}$}}
\newcommand{\secs}{\mbox{$\mbox{s}$}}
\newcommand{\ns}{\mbox{$\mbox{ns}$}}
\newcommand{\ms}{\mbox{$\mbox{ms}$}}
\newcommand{\mn}{\mbox{$\mbox{mn}$}}
\newcommand{\nm}{\mbox{$\mbox{nm}$}}
\newcommand{\mm}{\mbox{$\mbox{mm}$}}
\newcommand{\Hz}{\mbox{$\mbox{Hz}$}}
\newcommand{\days}{\mbox{$\mbox{day}$}}
\newcommand{\kHz}{\mbox{$\mbox{kHz}$}}

\begin{document}

\begin{frontmatter}

\title{Counting photons at low temperature\\with a streaming time-to-digital converter
}

\author[add_queens]{P.~C.~F.~Di~Stefano}
\ead{distefan@queensu.ca}
\author[add_queens]{P.~Nadeau}
\author[add_kvi]{C.~J.~G.~Onderwater}
\author[add_queens]{C.~Trudeau}
\author[add_queens]{M.-A.~Verdier}
\address[add_queens]{\queens}
\address[add_kvi]{\kvi}

\date{\today}

\begin{abstract}
We present some aspects of  photon counting to study scintillators at low temperatures.  A time-to-digital converter (TDC) had been configured to acquire several-minute-long streams of data, simplifying the multiple photon counting coincidence  technique. Results in terms of light yield and time structure of a \zwo\ scintillator are comparable to those obtained with a fast digitizer.  Streaming data also provides flexibility in analyzing the data, in terms of coincidence window between the channels, and acquisition window of individual channels.  We discuss the effect of changing these parameters, and use them to confirm low-energy features in the spectra of the number of  detected photons,
such as the $60~\keV$ line from \am\ in the \zwo\ sample.
We lastly use the TDC to study the transmission of the optical cryostat employed in these studies at various temperatures.
\end{abstract}


\end{frontmatter}

\linenumbers

\section{Introduction}
Rare-event searches such as those for neutrinoless double beta decay or for  dark matter have fueled recent interest in developing cryogenic scintillators for particle detection~\cite{stefano_foreword_2009,mikhailik_performance_2010}.  For a given energy deposited by a particle in a scintillator, the amount of light emitted depends on the nature of the particle.  When coupled to the measurement of phonons that is possible at low temperature (generally below $100~\mK$) and that provides the deposited energy, the measurement of scintillation photons therefore allows particle identification, and rejection of a significant fraction of the radioactive background in rare-event searches~\cite{lang_search_2009}.  The wealth of known room-temperature scintillators~\cite{rodnyi_physical_1997} also motivates the study of scintillators at low-temperature.

A typical method to study scintillators at low temperatures involves an optical cryostat with photomultipliers (PMs) at room temperature.
When a particle interacts in the cooled scintillator sample, the latter emits 
light
that can escape the cryostat and be detected  by the PMs~\cite{verdier_2.8_2009}.
One useful method is the multiple photon counting coincidence  (MPCC) technique~\cite{kraus_multiple_2005}.
Its hardware component uses the scintillation-induced coincidence between two PMs observing the crystal to start digitizing the  signals of each PM.  
Requiring a coincidence between the PMs can greatly reduce the number of recorded background events (from dark currents, for instance), though spurious coincidences from backgrounds can remain.
Offline, software extracts the arrival time and other information for each photon found in the trace.
The light yield is then determined from the number of photons, and the pulse shape is determined from the photon arrival times.
Compared to other methods like the delayed-coincidence technique~\cite{bollinger_measurement_1961}, the MPCC technique is well suited to the long time constants occurring at low temperatures, often greater than $100~\micros$, and supplies information on the number of photons emitted (and thus the light yield of the scintillator) in addition to just timing information.  
However, the quantity of acquired data is voluminous,
as is the time required to process it.  
Moreover, the result of the procedure (a list of photon times) is similar to what one would obtain from a single-start-multi-stop time-to-digital converter (TDC); this motivates the use of an actual TDC instead.  Indeed, multi-stop TDCs have already been used to increase the acquisition rate of the delayed-coincidence technique~\cite{moses_method_1993}.
In addition, a TDC may be able to write long streams of photon arrival times for each PM, thereby allowing flexibility in choosing the coincidence and acquisition windows offline, rather than having to set them ahead of time as is the case with the standard setup.

In this paper, we present the TDC we have used for this variant of the MPCC method, and compare its results to the standard, digitizer based, method.  We discuss optimization of the coincidence and acquisition parameters, and their influence on the low energy spectrum obtained with a \zwo\ crystal and \am\ radioactive source.  Lastly, we use the TDC to also characterize the transmission of our optical cryostat

\section{Running the TDC in streaming mode}
\label{sec_stream}
The standard MPCC method~\cite{kraus_multiple_2005} allows simultaneous measurement of light yield and time structure of scintillators.  It is well adapted to the long time constants (of the order of $100~\micros$ or greater) that can be encountered at low temperatures, and relies on a hardware component and a software one.  It requires a hardware trigger (for instance the coincidence between two photomultipliers, that can be as long as $1~\micros$ or greater) to start a digitizer (i.e. analog to digital converter or ADC) reading one or more photomultiplier tubes for a fixed duration.  A whole pulse trace is digitized each time, with a short enough sampling (typically between $1~\ns$ and $20~\ns$ depending on the PMs and preamplifiers) to resolve individual photoelectrons.  Offline, software identifies individual photons in each pulse, and extracts mainly the arrival time of each photon.  
Software cuts are then applied to remove spurious events from the set, and build spectra and pulse shapes.  The hardware and photon identification algorithms function in essence like a single-start-multi-stop time-to-digital converter (TDC); however, the quantity of data generated by the digitizer may be quite large ($10000$ events digitized on two channels with a sampling of $1~\ns$ for a duration of $1~\ms$ at a resolution of $8$~bits amounts to $20~\Gb$ of data\footnote{It's a testament to computing progress that just as in $1993$, when data sizes of $0.5~\Mb$ were deemed nearly prohibitive in this field~\cite{moses_method_1993}, the data volumes described here may be considered trivial in the future.}), and the time required to process it also, hence the motivation to use an actual TDC that would read the photomultiplier tubes directly.  
One possibility would be to use a hardware coincidence between two photomultipliers to start the TDC. However, since most TDCs  do not possess a pretrigger memory, 
this would lead to the loss of all early photons   before the second photon which determines the coincidence.  
In addition, it is not practical to delay the TDC signal by the duration of the coincidence window (often greater than $1~\micros$) or more because of signal attenuation.

\begin{figure}[hp]
	\centering
	\epsfig{file=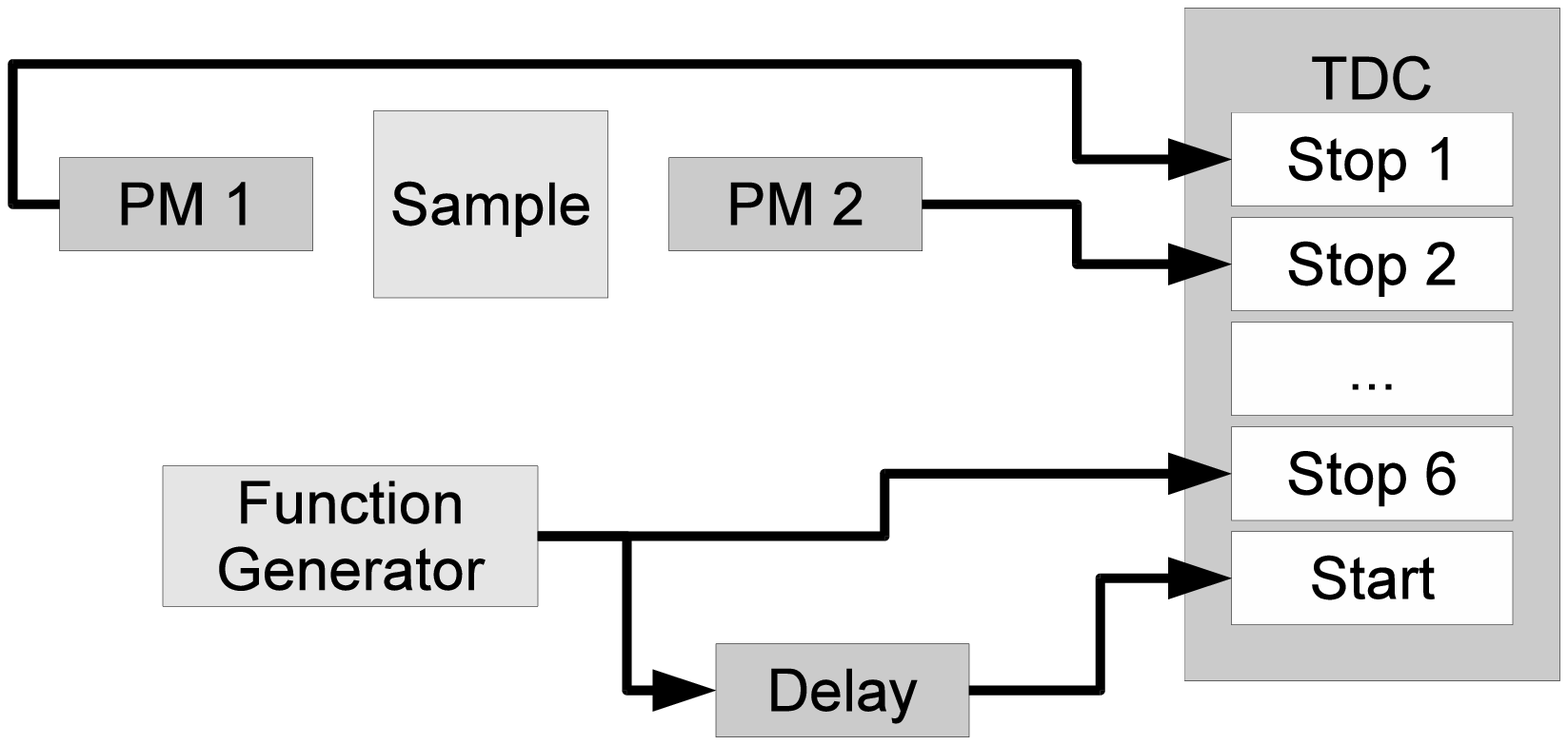,width=\linewidth}
	\caption[TDC setup]{Setup used to obtain streams of data from the TDC as applied to the MPCC technique.  A delayed signal from a periodic function generator starts the TDC.  The direct signal from the function generator, as well as the two signals from the PMs, are fed into the stop channels of the TDC.  To precisely measure the delay (created by a long wire), the direct and delayed lines from the function generator are swapped in a dedicated measurement.  The optical cryostat has been omitted for clarity.}
	\label{fig_tdc_setup}
\end{figure}

We have therefore turned our attention to running a TDC in continuous, streaming, mode.  In this configuration, the TDC is started at an arbitrary time, and then acquires stop signals (hereafter simply referred to as stops) nearly continuously during a run, i.e. a duration of the order of an hour broken into segments of several minutes.
Though TDCs able to run in this mode exist off-the-shelf (e.g. FAST~ComTec Gmbh MCS6A), we have access to a Compact PCI Agilent~U1051A (TC890) that requires a slightly modified configuration~\cite{_user_2010,meister_u1051a_2010}.  In standard operation, following a start signal on a dedicated channel, stop signals on up to six channels are recorded up to the next start signal or for $10.48~\ms$ at most.
Nominal retriggering time of each channel is $15~\ns$, and nominal time resolution on individual triggers of $0.05~\ns$.  
The device in fact has two memory banks that can operate alternately, allowing continuous acquisition of stops over longer periods of time, for instance by using a repetitive start function with a period smaller than $10.48~\ms$ --- provided the rate of events is reasonable.
To obtain a continuous stream, we triggered the TDC with a National Instruments PXI~5422 function generator running at $110~\Hz$, 
in view of concatenating the resulting segments into a nearly continuous stream, as illustrated in Figure~\ref{fig_tdc_setup}.  
The TDC records stop times relative to the nearest start, so reconstructing a stream of stops much longer than $10.48~\ms$ requires knowing the start times precisely.
This would be straightforward if the $9.1~\ms$ period of the function generator was stable and known to a better precision than the $0.05~\ns$ precision of the stops, which unfortunately may not be the case.
To circumvent this, the signal from the start generator was split into a direct signal which was sent to one channel of the TDC, and a delayed signal used to start the TDC.  The time of the next start relative to a given one is therefore measured for each event, provided the delay is known precisely. The delay was produced by a length of cable, and was measured as $61~\ns$ in a dedicated run in which the direct signal and delayed signals were swapped: the former started the TDC, and the latter was sent to one of the stop channels.  
To facilitate data management, the LabView DAQ controlling the TDC breaks the data into files typically corresponding to  some five minutes in length.
Software written in Java was then used to reconstruct the streams in a given file.  The software interrupts a stream any time memory errors are encountered. We have found that the following TDC acquisition parameters ensure errors are few: $100$ starts per memory transfer and $360$ memory transfers per file (i.e. streams of $327~\secs$ per file)

To test the efficiency of our reconstruction, we ran this setup for some $5$~minutes using  a $110~\Hz$  ($T =9.1~\ms$) frequency square start signal, and used as stops a square signal coming from another function generator with a frequency of roughly $73~\Hz$ ($T'=13.7~\ms$), then studied the differences between reconstructed stop times.  The direct times were also sent to one of the stop channels.  Let $T$ be the true start period, and $T<T'<2T$ be the true stop period.  The starts arrive at times $t_i=iT$, and the stops at times $t'_j=\varepsilon+jT'$, where $0 \leq \varepsilon < T$ represents the arbitrary phase shift between the signals.  The TDC in fact measures $\delta_j$, the arrival time of stop $j$ after its preceding start; reconstruction assumes that times $t_i$ are known.  Given the periods used, each start is followed by one or zero stops.  The phase shift $\varepsilon$ is constant between gaps (starts with no stops).  Reconstruction of stops not separated by gaps is done by $T'=\delta_{i+1} - \delta_{i} + T$; for those separated by a gap, $T'=\delta_{i+1} - \delta_{i} + 2T$.  If a start was followed by a stop, and both are lost because of some memory problem, then reconstruction will incorrectly overestimate the next stop times and the period $T'$ will be overestimated as $\delta_{i+2} - \delta_{i} + T = T' + \Delta T$, where $\Delta T \equiv T'-T$.  If a gap was lost, then $T'$ will be underestimated as $\delta_{i+2} - \delta_{i} + T =\Delta T$.  If only a stop is missed, then the period will be overestimated as $2T'$.

\begin{figure}[hp]
	\centering
	\epsfig{file=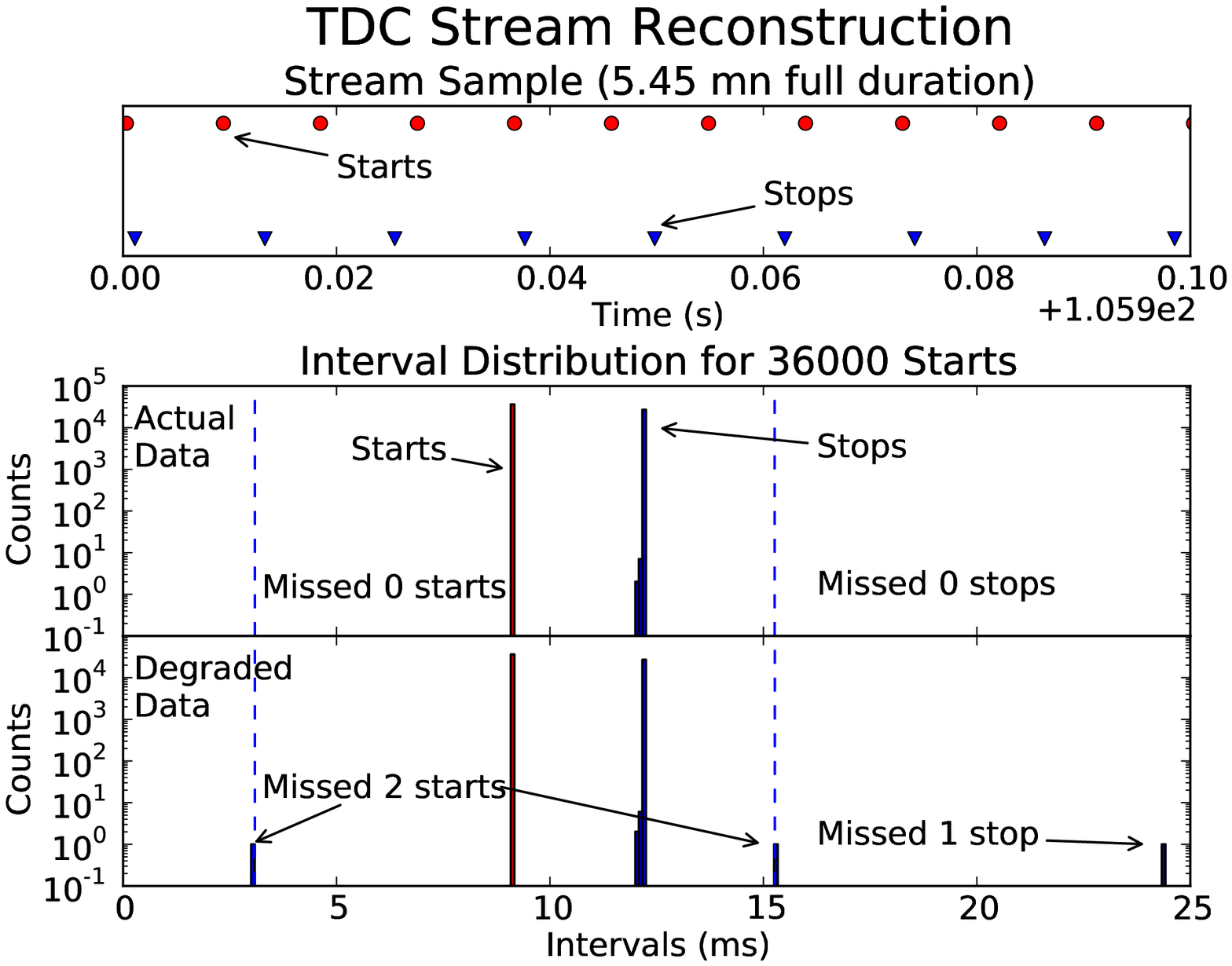,width=1.0\linewidth}
	\caption[TDC tests]{Top: reconstructed pattern of TDC $36000$~starts and stops. Full stream lasts roughly $5~\mn$.  Starts and stops have different periods.  Middle: intervals calculated from above streams.  No events are observed outside of the dashed blue lines, indicating no starts or stops have been missed during reconstruction.  Bottom: same as middle, but with data that have been voluntarily degraded to evidence two types of missed starts and one type of missed stop.}
	\label{fig_tdc_tests}
\end{figure}

Fig.~\ref{fig_tdc_tests} shows examples of reconstructed start and stop times, as well as the histogram of the difference between consecutive start times and stop times.  The mean values of these histograms determine $T$ and $T'$ respectively. The histogram of stop time differences is free of entries 
at or below $\Delta T$ and at or above $T' + \Delta T$
implying no starts or stops are missed and that the stream has been properly reconstructed over this $\approx 5~\mn$ interval. 
As a control, we have manually degraded the same data set, removing one start and its stop, removing a start that is not followed by a stop, and removing a stop.  As expected, these respectively induce misreconstructed periods at $T' + \Delta T$, at $\Delta T$, and at $2T'$.
As an additional  consistency check, we have calculated the number of expected stops based on the ratio of measured start to stop periods and the number of observed starts.  This yields an expected number of stops of $26873.8$, in excellent agreement with the measured number of stops, $26873$.
We note that the reconstructed stop periods distribution (average $12.18~\micros$) is asymmetric with a few periods that are shorter than expected (of the $36000$ events, the shortest is $12.06~\micros$).  The standard deviation of  $2~\micros$ for a period of $12~\ms$,  is small, but greater than the corresponding numbers for the start period ($7~\ns$ for $9.1~\ms$).  We are unable to conclude if this is caused by our technique or instabilities in the function generator.  

\begin{figure}[hp]
	\centering
	\epsfig{file=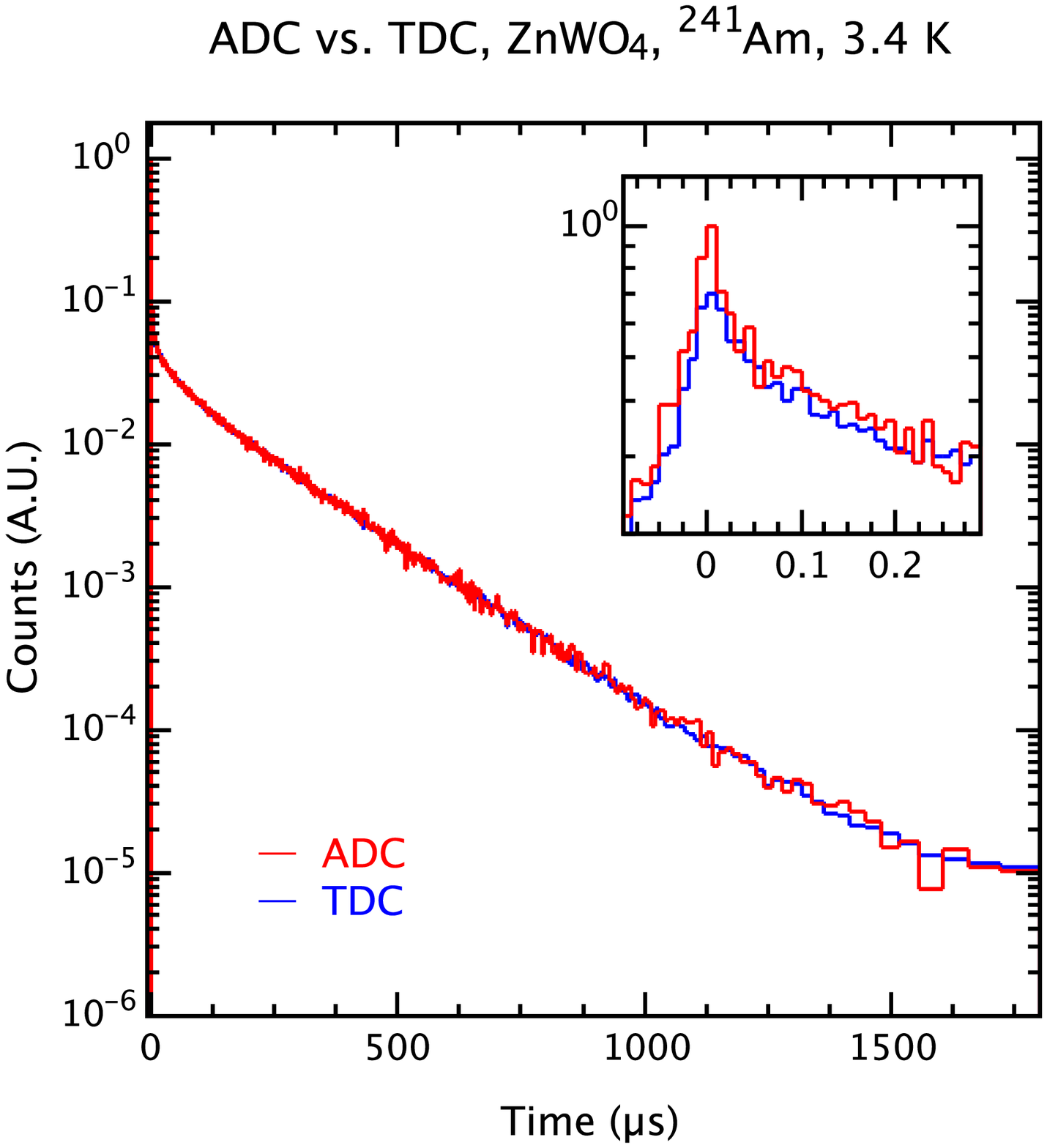,width=\linewidth}
	\caption[TDC vs PXI]{Comparison of pulse shapes obtained with the MPCC method using either the TDC or a digitizer (ADC).  Pulses have been normalized so that the main decays overlap.  The shapes are identical except for the shortest times, since the deadtime between TDC stops is greater than that obtained with the digitizer.}
	\label{fig_tdc_pxi}
\end{figure}

In practice, the consequences of these small imperfections are unobserved, in part because if events are reconstructed over $1~\ms$ for instance, then only one in nine events will overlap consecutive starts.
Once software has reconstructed the streams, a second routine identifies coincidences and outputs the data in our usual MPCC format, so that our standard MPCC analysis routines can be used.  
We have tested our analysis pipeline from the identification of coincidences and on by generating simulated streams of data discussed in Sections ~\ref{sec:CoincWindow} and~\ref{sec:AcqWindow}.
Another test we have carried out is the comparison of results from the TDC and from the standard digitizer-based MPCC method with a National Instruments PXI~5154 digitizer used previously~\cite{verdier_scintillation_2011}.  
These results, in terms of average pulse shape, were obtained with a $20\times 10 \times 5~\mm^3$ \zwo\ crystal  at $3.4~\kelv$ and are shown in Figure~\ref{fig_tdc_pxi}.
The sample was provided by the CRESST collaboration, and \zwo\ is being actively considered as a scintillator for dark matter searches~\cite{kraus_multiple_2005,bavykina_development_2009,kraus_znwo4_2009}. 
The figure shows that the two methods yield very similar results except at short times of the order of a few tens of nanoseconds, where the TDC underestimates the number of photons compared to the digitizer.  This is consistent with the TDC not being able to resolve stops closer than $15~\ns$ apart, whereas the digitizer and the offline photon identification routines can resolve individual photons separated by as little as $5~\ns$~\cite{verdier_scintillation_2011}.
Other differences between the digitizer and the TDC include the fact that with the former, it is possible to obtain the amplitude, or the integral, of each photon pulse, and it is possible to reanalyze data offline changing the threshold for the photon, whereas the TDC allows reanalysis of the data offline using different coincidence and acquisition windows.
All in all, the TDC is a viable alternative to a digitizer for the MPCC technique, except perhaps at short time constants.  It should be noted however that at very short time constants, the MPCC method itself is not recommended~\cite{kraus_multiple_2005}.

\section{Choice of the coincidence window}
\label{sec:CoincWindow}
\begin{figure}[hp]
	\centering
	\epsfig{file=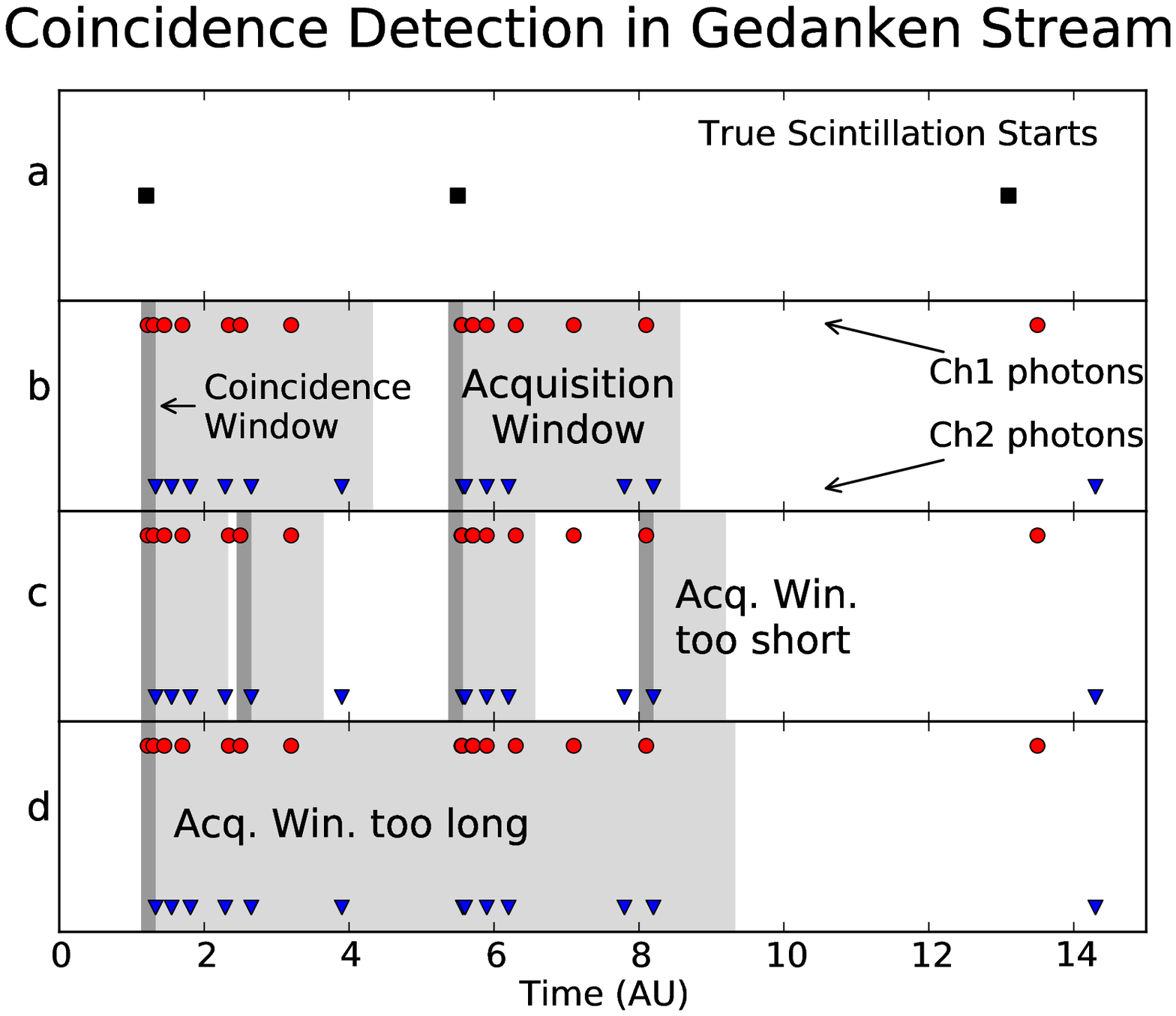,width=\linewidth}
	\caption[Coincidence detection]{Sketch of detection of events in  reconstructed stream. a)~Actual start of random scintillation events, not directly measured. b)~Arrival time of photons on both channels, and illustration of coincidence windows (dark gray) and acquisition windows (light gray). c)~Example of an acquisition window that is too short, leading to multiple coincidence and triggers on a same scintillation event. d)~Example of an acquisition window that is too long, leading to overestimation of photons in a given event.  Pretrigger region has been omitted for clarity.}
	\label{fig_coinc_stream}
\end{figure}

Once continuous streams of photons have been obtained from the TDC for both PMs, a software algorithm looks for coincidences between the two channels, and, when one is found, identifies the times of photons for a given acquisition window, as illustrated in Figure~\ref{fig_coinc_stream}. 
Unlike the digitizer based approach, in which both coincidence window ($T_{coinc}$) and acquisition window ($T_{acq}$) are set once and for all, with the TDC, these parameters can be adjusted after data have been taken.  This is convenient since the scintillation time constant of the sample being studied is not necessarily known in advance, yet both parameters are related to it. We also note that in the standard hardware coincidence setup, the time of the second of the two photons involved determines the coincidence, whereas by software, the start of the coincidence can be chosen as either of the two photons.  In both the hardware and software coincidence techniques, data are also recorded over a pretrigger usually chosen to include the coincidence window.

We first consider the influence of the coincidence window on the shape of the spectra, with no cuts applied to the data.  For a given coincidence window $T_{coinc}$, and a scintillator emitting uncorrelated photons with an exponential time constant $\tau$ such that $n$ are detected by one PM and $m$ by the other, the coincidence probability is:
\begin{equation}
\label{eq:Simple_Coinc_mn}
p_{\mbox{coinc}} =  1- e^{-nmT_{coinc}/\tau}
\end{equation}
This expression, derived  in \ref{app:CoincProba} and generalized there to multiple time constants (Eq.~\ref{eq:MultiCoincmn}), depends on the dimensionless number $nmT_{coinc}/\tau$ which is the product of the number of photons and the coincidence window over the time constant.  For a luminous scintillator with a short time constant or a long coincidence window, this number, and therefore the coincidence, probability is high.  Conversely, for few photons, or a slow scintillator, or a short coincidence window, the number and the coincidence probability is low.  One consequence of this is that for a given scintillator, coincidence window, and photons detection efficiency, the  coincidence efficiency is lower for the low energy part of the energy spectrum since there are fewer photons.  This can distort the shape of the spectrum, mainly in terms of amplitudes of various lines, but could also shift some lines to a small extent.  This last effect would be relevant when studying the linearity of a scintillator or its quenching factor for various particles.  Correcting the photon spectra, which show the histograms of the sum of photons on both channels (i.e. $n+m$), for coincidence efficiency can be attempted by two approaches.  In the bin-by-bin approach, one assumes that $n=m=\frac{n+m}{2}$ and corrects the spectrum by dividing  each bin  by the function $1- e^{-\left( \frac{n+m}{2} \right)^2T_{coinc}/\tau}$.  In the  more precise event-by-event approach, as each event is binned into the histogram, it is weighted by the inverse of $1- e^{-nmT_{coinc}/\tau}$.  The difference between the two methods is rather small in our case since the optical efficiencies of the PMTs are similar, with the  exception of cases in which the number of photons is low and statistical fluctuations become important.

\begin{figure}[hp]
	\centering
	\epsfig{file=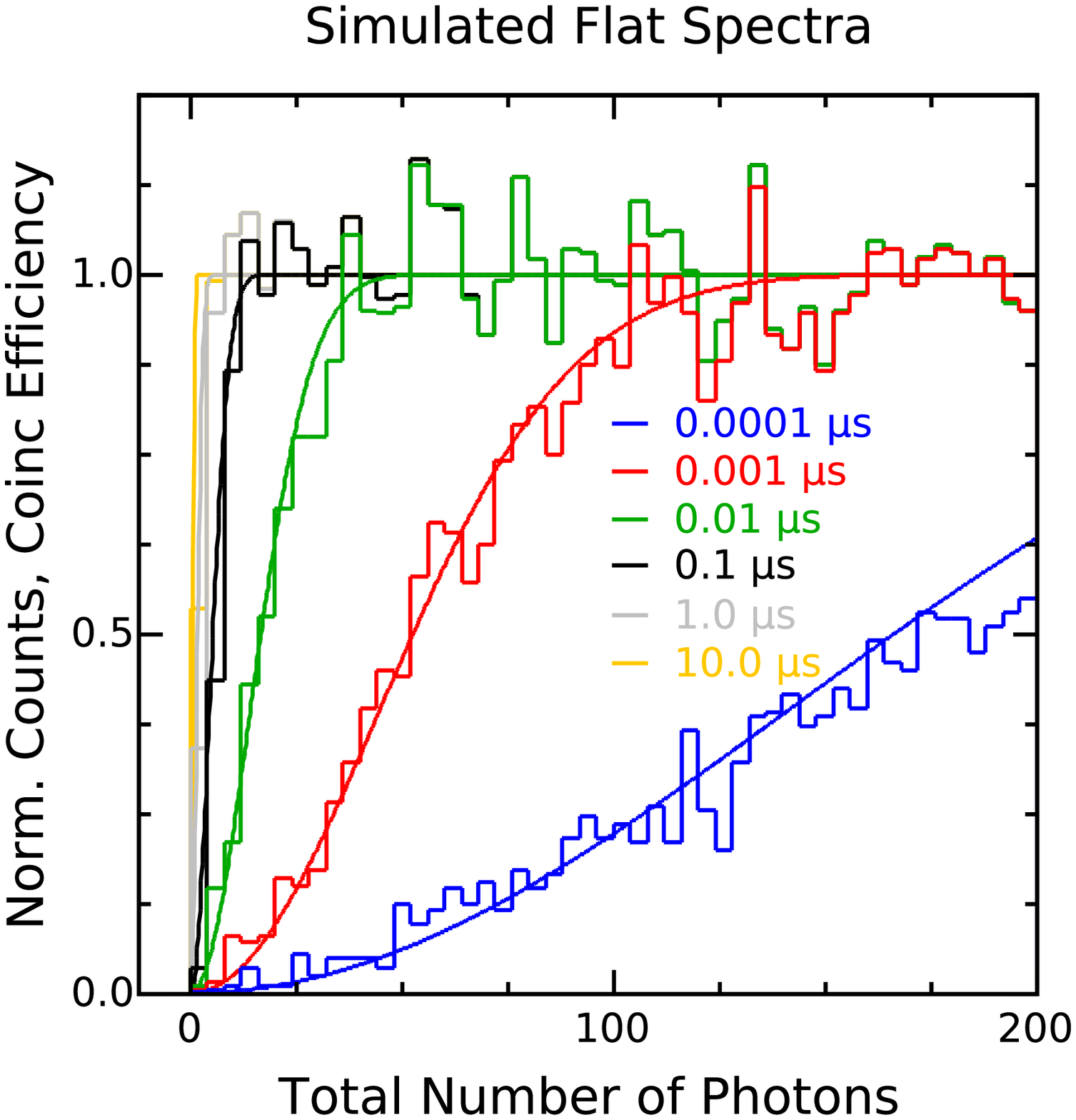,width=0.9\linewidth}
	\caption[Trigger efficiency]{Comparison of calculated coincidence efficiencies and a simulated flat spectrum processed by the analysis pipeline, for various coincidence windows. Time constant of the scintillator used in simulation is  $1~\micros$; for each event, the number of photons on each channel are drawn from a Poisson distribution whose expected value is itself drawn from a uniform distribution with maximum $100$. Abscissae in the plot are the sum of photons on both channels. Spectra are normalized to number of events in the simulation.  For a given coincidence window, as the number of photons increases, so does the probability of coincidence.  For a given number of photons, as the coincidence window increases, so does the probability of coincidence.  The effect is well represented by Eq.~\ref{eq:Simple_Coinc_mn}.  For clarity in this figure and the next, error bars are omitted but can be estimated from the bin-to-bin fluctuations.}
	\label{fig_trig_eff_sim}
\end{figure}

In Fig.~\ref{fig_trig_eff_sim}, we compare the theoretical coincidence efficiencies obtained from Equation~\ref{eq:Simple_Coinc_mn}  with simulated data that have been processed by the analysis pipeline.    
A data stream was generated for $10^4$ events coming from a scintillator of time constant $1~\micros$.  To avoid pileup, the full stream is assumed to last $10^8~\micros$.
The expected number of photons generated for each event is itself drawn from a  distribution that is flat with a maximum of $100$ photons per channel.  For each event, and each channel, there are Poisson fluctuations around the expected number of photons.
The full set of photons is sorted by arrival time and written to a stream file.
These simulated data are then fed into the analysis pipeline with various coincidence windows logarithmically spaced from $10^{-4}~\micros$ to $10^{2}~\micros$.  The analysis uses identical acquisition windows of $9~\micros$ ensuring that  more than $99.9\%$ of photons should be counted. The  $1~\micros$ preceding the coincidence is used for pretrigger.
Figure~\ref{fig_trig_eff_sim} shows that the flat spectra are suppressed for low numbers of photons and small coincidence windows, and that the effect is well reproduced by Equation~\ref{eq:Simple_Coinc_mn}.

\begin{figure}[hp]
	\centering
	\epsfig{file=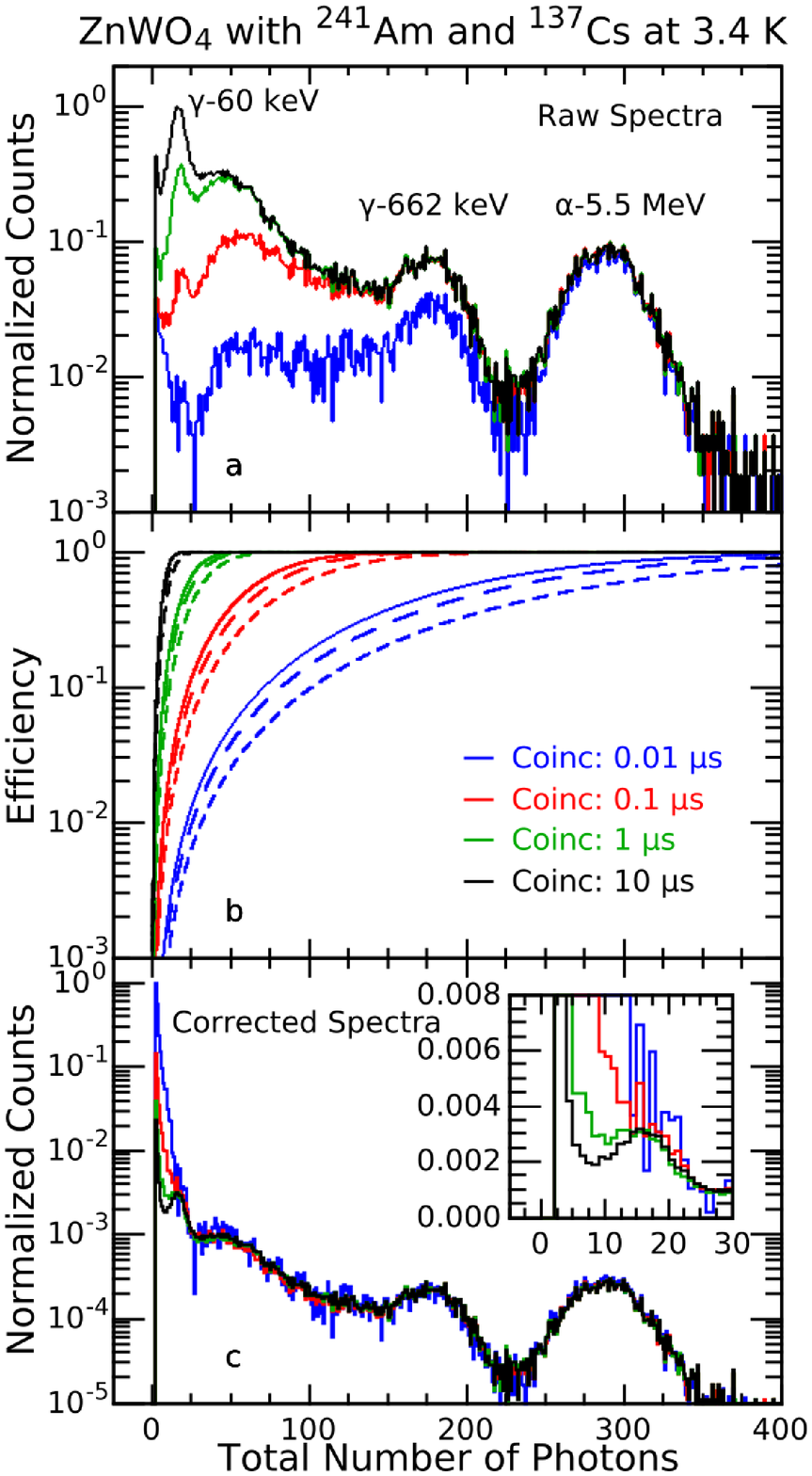,width=1.0\linewidth}
	\caption[Trigger efficiency]{Effect of coincidence window length on spectra obtained from a \zwo\ crystal exposed to $\alpha$ particles from a \am\ source and to $\gamma$ particles from a \cs\ source, at a temperature of $3.4$~\kelv.  Top: raw spectra, for various coincidence windows.  Middle: threshold efficiency functions derived from Eq.~\ref{eq:MultiCoincmn}.  Calling $R$ the ratio of photons on each channel, the solid curves are for $R=1$, the long dashed ones for $R=3$, and the short dashed ones for $R=6$. Bottom: spectra corrected for threshold efficiency using the event-by-event method.  Despite the correction, the low energy portion of the spectrum is not properly resolved if the coincidence window was too short (see insert with linear ordinates).
	}
	\label{fig_trig_eff_data}
\end{figure}

We have carried out a similar analysis using real data obtained from the CRESST \zwo\ crystal at $3.4~\kelv$ in Fig.~\ref{fig_trig_eff_data}. 
The sample was exposed to $\alpha$ and $\gamma$ particles from an \am\ source, and concurrently to $\gamma$ particles from a \cs\ source.  Data were analyzed using a fixed acquisition window of $1.8~\ms$, a pretrigger of $0.2~\ms$, and coincidence windows ranging from $0.01~\micros$ to $10~\micros$.
The top figure shows the rough, uncorrected spectra.  The $\alpha$ peak, around $300$~photons summed on both channels, is the same in all cases, but the lower $662~\keV$ \cs\ line around $180$~photons already shows some distortion, and there is more than an order of magnitude difference in the spectra around $50$~photons.  The larger coincidence windows also show some structure around $18$~photons which is absent from the shorter coincidence windows.  The position of this structure is consistent with it being the $60~\keV$ line from \am.
The middle figure shows the efficiencies for different coincidence windows, assuming the time constants and numbers of photons obtained for \zwo\ and $\gamma$ particles in a double-coincidence setup triggered $511~\keV$ photons from a \ttna\ source~\cite{verdier_scintillation_2011}.  Curves assume different number of photons on both channels, parametrized by $R$, the ratio of photons on each channel.
The bottom figure shows the spectra corrected for the efficiency curves.  The corrected spectra are in good agreement with one another down to at least $50$~photons; i.e. the correction works for coincidence efficiencies at least as low as $10\%$.  
The spectra are now compatible around the $60~\keV$ line in terms of position and amplitude, though the line is not resolved for the shorter windows.
The positions and amplitudes of the $60~\keV$ line match for the longer coincidence windows.  
This buttresses our earlier identification of this line~\cite{verdier_setup_2012}.
The spectra from the shorter coincidence windows are compatible with the presence of the same line though it can not be resolved from the noise.  
This shows that data taken using the standard MPCC technique with a poor choice of hardware coincidence window can not necessarily be corrected perfectly at low number of photons.
Conversely, the flexibility afforded by the streaming TDC  to adjust the coincidence window after data have been taken assists in understanding  the coincidence efficiency and helps to ascertain the low energy features of the spectra.

\section{Choice of the acquisition window}
\label{sec:AcqWindow}

The next parameter we study is the duration of the acquisition window ($T_{acq}$) over which photons are recorded.
In the MPCC technique, the average time of pulses, defined as the average arrival time of photons after the first one, is used to reject events suffering from pileup~\cite{kraus_multiple_2005}.  However,  in certain conditions, the average time of the pulse may itself be biased by pileup.  For photons distributed according to several exponential distributions, the mean arrival time of photons counted from the start of the pulse over a time $T_{acq}$ is
\begin{eqnarray}
\label{eq:AvTime}
\frac{ \int_0^{T_{acq}} t \frac{dn}{dt} dt}{\int_0^{T_{acq}} \frac{dn}{dt} dt}  & = & \frac{\sum n_i \tau_i \left(  1 - e^{-T_{acq}/\tau_i} \left(1+  T_{acq}/\tau_i \right) \right)}{\sum n_i \left(  1 - e^{-T_{acq}/\tau_i} \right)} \nonumber \\ & \leq  & \frac{\sum n_i \tau_i}{\sum n_i}
\end{eqnarray}
(\ref{app_avtime_photons}).
For an infinite window, this yields $\frac{\sum n_i \tau_i}{\sum n_i}$, referred to here as the effective time constant.  For a shorter window, the value will be underestimated.  For long values, pile-up will lead to an overestimation of this parameter.

Once the average arrival time is determined, cuts based on the time of the first photon and on the average time can be applied to the data to reject pileup in certain cases, and the light yield can be studied with histograms of the number of photons.
Using the same notations as before,  the number of photons actually counted during acquisition window $T_{acq}$ if there is no pileup will be:
\begin{equation}
\label{eq:NumPhots}
\int_0^{T_{acq}} \frac{dn}{dt} dt = \sum_i n_i \left( 1 - e^{-T_{acq}/\tau_i}\right) \leq \sum n_i
\end{equation}
For an infinite window, this number is $\sum n_i$.  If the acquisition window is too short, then photons will be missed.  Practically, $90\%$ of the photons can be counted for $T_{acq}/\tau = 2.3$.
Another effect  can occur when the acquisition window is too short compared to the time constant of the scintillator (Figure~\ref{fig_coinc_stream}c):  after a first coincidence is detected at the start of an event, the acquisition window is too short to cover the length of the pulse, and a second coincidence can be detected right after the acquisition window on the remaining photons.  
This second acquisition window will only contain a fraction of the photons which can appear as artefacts in the spectrum of the number of photons. 
This effect can be repeated again and again, introducing spurious features in the spectra with the following, decreasing, number of photons:
\begin{equation}
\label{eq:NumPhotsArtefacts}
\nu_j \equiv \int_{jT_{acq}}^{(j+1)T_{acq}} \frac{dn}{d\tau} dt =  \sum_i n_i e^{-jT_{acq}/\tau_i} \left( 1 - e^{-T_{acq}/\tau_i}\right)
\end{equation}
For example, a single time constant of $\tau=165~\micros$ and an acquisition window of $T_{acq}=200~\micros$ result in $\nu_0/n=0.70$, $\nu_1/n=0.21$ and $\nu_2/n=0.06$.  We do not attempt to calculate the number of events for each $\nu_j$, though this may be possible by a reasoning on the coincidence probability of the remaining photons along the lines of that in~\ref{app:CoincProba}.

Moreover, if the acquisition window is too long, then there is a risk that pileup will occur. For instance, if the acquisition window is long compared to the average time between events, the spectrum of number of photons will be biased (Figure~\ref{fig_coinc_stream}d).  If the acquisition window also happens to be larger than the scintillation time constant, several events will be fully contained in the window, and the true number of photons (essentially $\sum n_i$) will give rise to spurious artefacts at integer multiples $j \sum n_i$.

\begin{figure}[hp]
	\centering
	\epsfig{file=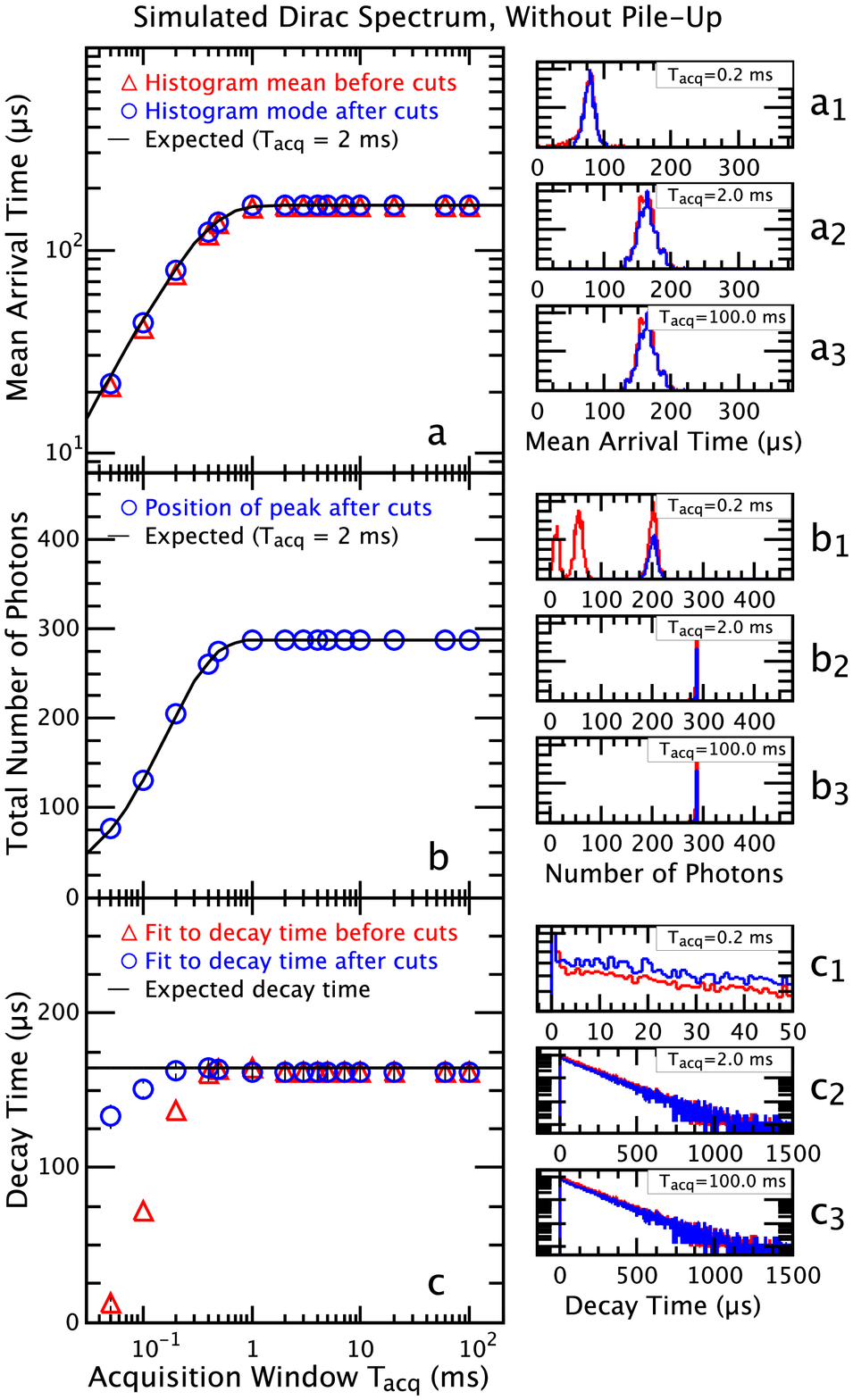,width=1.0\linewidth}
	\caption[Effect of acquisition window length on average time of event]{Results from $1000$ simulated events drawn over $2\times 10^{14}~\micros$ with $144$ photons per event per channel and a single time constant of $165~\micros$.
	Top: mean arrival time.  Middle: total number of photons.  Bottom: time constant determined from fit of average pulse.  
	In all plots, green is before cuts, blue after.  One standard deviation error bars are shown, but are often smaller than marker size.
	See text for discussion.
	}
	\label{fig_sim_acqWin_NoPileup}
\end{figure}

\begin{figure}[hp]
	\centering
	\epsfig{file=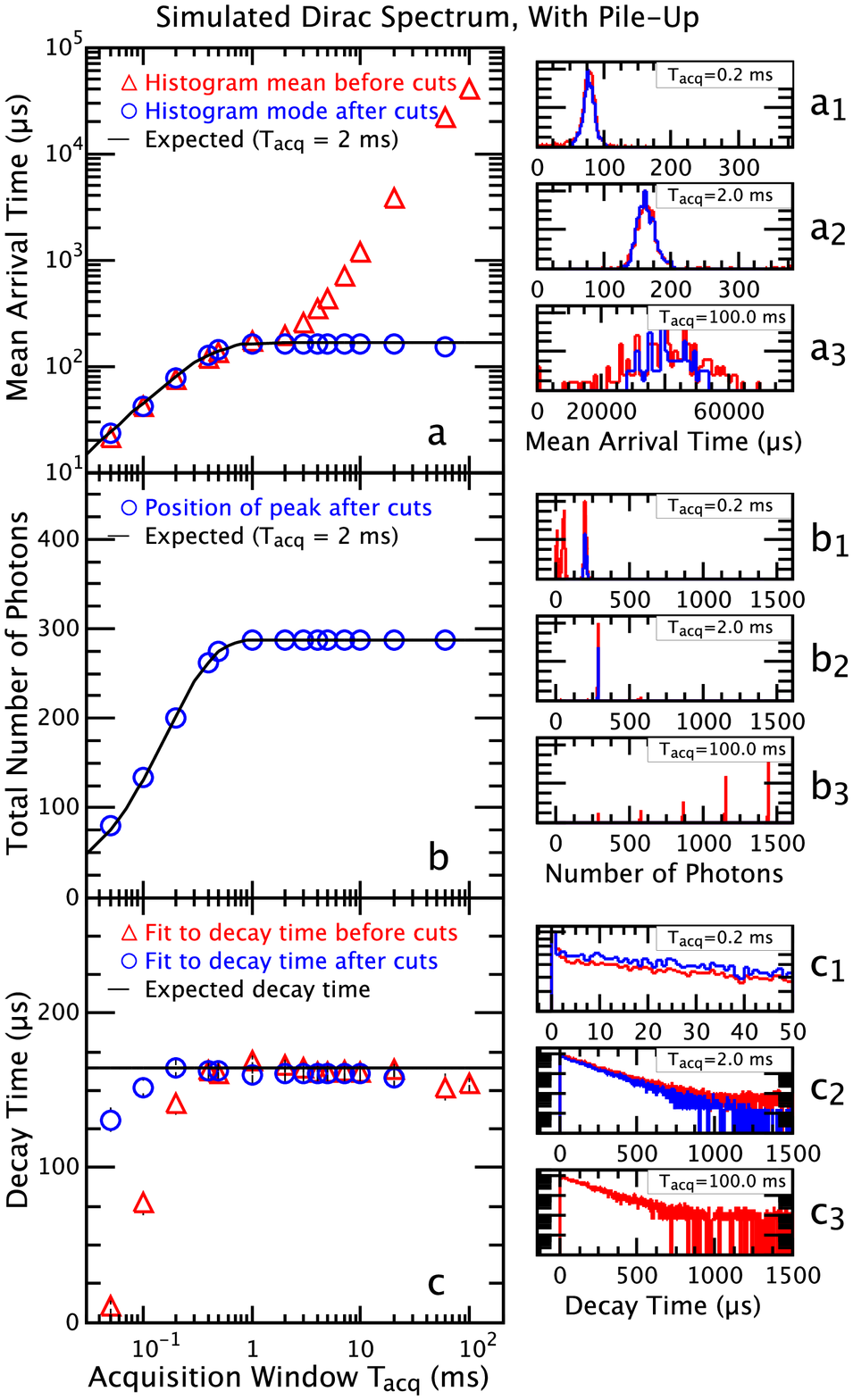,width=1.0\linewidth}
	\caption[Effect of acquisition window length on number of photons]{Simulation with same parameters as Fig.~\ref{fig_sim_acqWin_NoPileup}, but average time between events reduced to $2\times 10^{4}~\micros$ to simulate pileup. 
	}
	\label{fig_sim_acqWin_BadPileup}	
\end{figure}

We have studied the effect of the acquisition window using simulated streams of data.  In these simulations, a given number of events (typically $10^3$) are assumed to arrive randomly according to a uniform distribution over a given amount of time, with an average time between events of $\Delta T$.  
For each event, the expected number of photons on each channel is fixed.  For each event, the actual number of photons is  set to the expected number (no Poisson fluctuations). Once all the events have been drawn, they are sorted by time and written to a stream file that is fed into the analysis pipeline.
Results coming from a simulation with a single time constant of $\tau=165~\micros$ and Dirac delta function spectra centered on $144$~photons per channel 
are shown in Figures~\ref{fig_sim_acqWin_NoPileup} ($\Delta T  = 2\times 10^{11}~\micros$, i.e. no pileup)
and~\ref{fig_sim_acqWin_BadPileup} ($\Delta T =  2\times 10^{4}~\micros$, much pileup).
The data are first analyzed with the following standard MPCC cuts~\cite{kraus_multiple_2005}: cut on first photon arrival time, test statistic cut.  Then,  the time structure of the events can be studied by building a histogram of the arrival times of the photons for events with a well-defined number of photons.

In the case of data with no pileup (Fig.~\ref{fig_sim_acqWin_NoPileup}), the mean arrival time and the total number of photons behave  as expected from Equations~\ref{eq:AvTime} and~\ref{eq:NumPhots}.  For the total number of photons, the mode of the distribution is more robust than the mean, and at very short window times (Fig.~\ref{fig_sim_acqWin_NoPileup}$b_1$), the distribution itself displays the artefacts described in Eq.~\ref{eq:NumPhotsArtefacts} that may complicate determination of the mode.  The fitted position of the three visible peaks is broadly consistent with what is expected from Eq.~\ref{eq:NumPhotsArtefacts}: $202.9 \pm 0.2$ photons   (compared to $288 \times 0.7 = 201.6$) , $57.6 \pm 0.2$ (compared to $288 \times 0.21 = 60.5$), and $13.4 \pm 0.3$ (compared to $288 \times 0.06 = 17.3$).
In addition, the reconstructed pulse shapes show an excess of events in the first time bin (Fig.~\ref{fig_sim_acqWin_NoPileup}$c_1$).  This is a consequence of the time of each photon in an event being calculated relative to the first photon, rather than to the true start time of the event which is unknown.  For small acquisition windows, this causes the result of an exponential fit to the histogram to underestimate the true time constant (the fit also includes a flat background).  For larger acquisition windows however, the first bin has less weight relative to the others, and the fitted time constant closely matches the input of the simulation. One experimental modification that circumvents this problem and allows more precise measurements of short scintillation times is to use a fast extra scintillator and a tagged source~\cite{verdier_scintillation_2011}. Overall, on this clean data set, the expected time constant is properly reconstructed and the cuts play only a small role.

In the case of data with significant pileup (Fig.~\ref{fig_sim_acqWin_BadPileup}), 
the  histogram of the number of photons (Fig.~\ref{fig_sim_acqWin_BadPileup}$b_1$)
displays artefacts at positions similar to those in the no-pileup case when the acquisition window is short.  
The mean arrival time starts to diverge from the model when the acquisition window becomes long enough for pileup to become significant.
For large acquisition windows (e.g. $T_{acq} = 100~\ms$, Fig.~\ref{fig_sim_acqWin_BadPileup}$b_3$), the distribution of the number of photons exhibits spurious peaks at multiples of the true number, as the window is large enough to include several events ($T_{acq} \gg \Delta T$), and the time constant of the events is also much smaller than the acquisition window ($T_{acq} \gg \tau$).  
When this is accounted for and the proper peak selected, the total number of photons follows the model.
After cuts, the proper pulse shape is reconstructed, except for the longest windows in which the prevalence of pileup leads to the cuts rejecting too many events for a fit on the pulse shape to be performed (Fig.~\ref{fig_sim_acqWin_BadPileup}$c_3$).

\begin{figure}[hp]
	\centering
	\epsfig{file=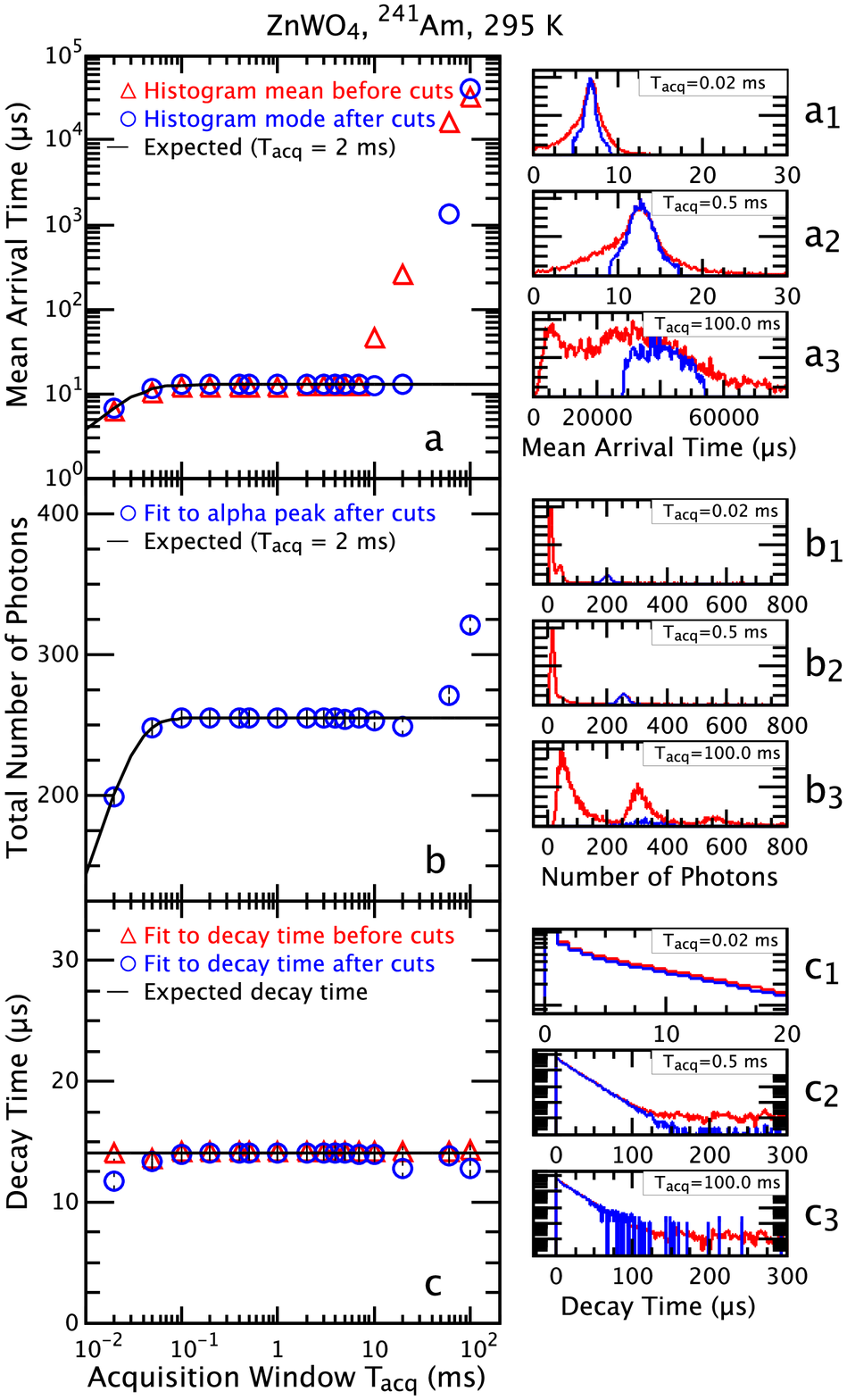,width=1.0\linewidth}
	\caption[Effect of acquisition window length on number of photons]{Same analysis as Fig.~\ref{fig_sim_acqWin_NoPileup}, but carried out on data from a \zwo\ crystal at $295~\kelv$ exposed to a \am\ source. Fits to pulse shape use $2$ exponentials and a constant --- only time constant with most photons is shown.  See text for details.}
	\label{fig_znwohot_acqWin}	
\end{figure}

\begin{figure}[hp]
	\centering
	\epsfig{file=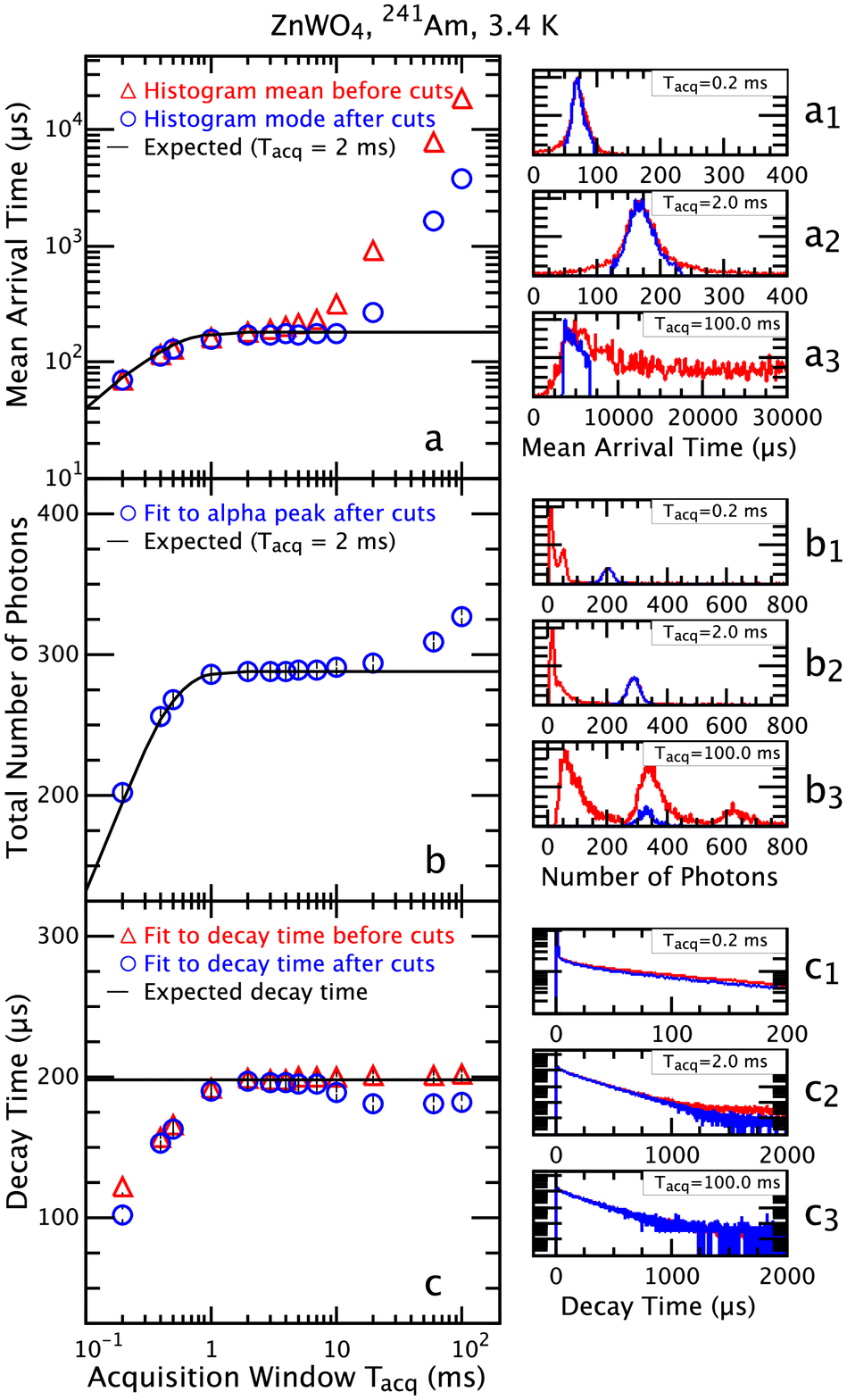,width=1.0\linewidth}
	\caption[Effect of acquisition window length on number of photons]{Same as Fig.~\ref{fig_znwohot_acqWin}, but \zwo\ crystal is now at $3.4~\kelv$, lengthening the time constants and increasing the light yield. Fits to pulse shape use $3$ exponentials and a constant --- only time constant with most photons is shown.   See text for details.}
	\label{fig_znwocold_acqWin}	
\end{figure}

A similar analysis is next carried out on \zwo, at two temperatures ($295~\kelv$ and $3.4~\kelv$) illustrated in  Fig.~\ref{fig_znwohot_acqWin} and~\ref{fig_znwocold_acqWin}.  
Unlike the simulation, the time constant of \zwo\ at various temperatures is not known {\em a priori}.
The analysis has been carried out with a $900~\ns$ coincidence window.
As the crystal cools, the effective time constant increases by an order of magnitude from $\approx~10~\micros$ to $\approx~200~\micros$. This increase is broadly consistent with fluorescence measurements~\cite{babin_decay_2004}.  At both temperatures, for short acquisition windows, the average time increases with the acquisition window (Fig.~\ref{fig_znwohot_acqWin}a and~\ref{fig_znwocold_acqWin}a).  
There is then a plateau during which the average time is independent of acquisition window. Up to here, there is good agreement with Eq.~\ref{eq:AvTime}.  However, when the window becomes too long, pileup appears and the average time increases once more, though the mode is slightly more robust.
From the standpoint of the number of photons, the short windows contain an artefact  echoing the main $\alpha$ line in the spectra, as per Eq.~\ref{eq:NumPhotsArtefacts} (Fig.~\ref{fig_znwohot_acqWin}$b_2$ and~\ref{fig_znwocold_acqWin}$b_2$).  In the long acquisition windows, there are also artefacts at integer multiples of the $\alpha$ line (Fig.~\ref{fig_znwohot_acqWin}$b_3$ and~\ref{fig_znwocold_acqWin}$b_3$).  As the acquisition window changes, the position of the $\alpha$ line itself follows the model in Eq.~\ref{eq:NumPhots} until pileup causes it to be overestimated (Fig.~\ref{fig_znwohot_acqWin}b and~\ref{fig_znwocold_acqWin}b).  The cuts allow the average pulse shape to be reconstructed in a way that does not depend very much on the acquisition window, except at low temperatures and short windows.  Pulses have been fitted using $2$ exponentials and a constant at $295~\kelv$, and using $3$ exponentials and a constant at low temperature.
Overall, since the time constants of the crystal change with temperature, an acquisition window that is correct at room temperature (e.g. $0.1~\ms$) may not be optimal at low temperature; and the TDC allows to make up for this after data have been taken.  This is important in particular to understand whether features in a spectrum are real (i.e. the $60~\keV$ line from \am) or artefacts (echoes of the $\alpha$ line from \am).  We defer the study of the differences in time constants between alpha and gamma particles to future work.

\section{Transmission of optical cryostat}
\label{sec:OpticaTrans}
\begin{figure}[hp]
	\centering
	\epsfig{file=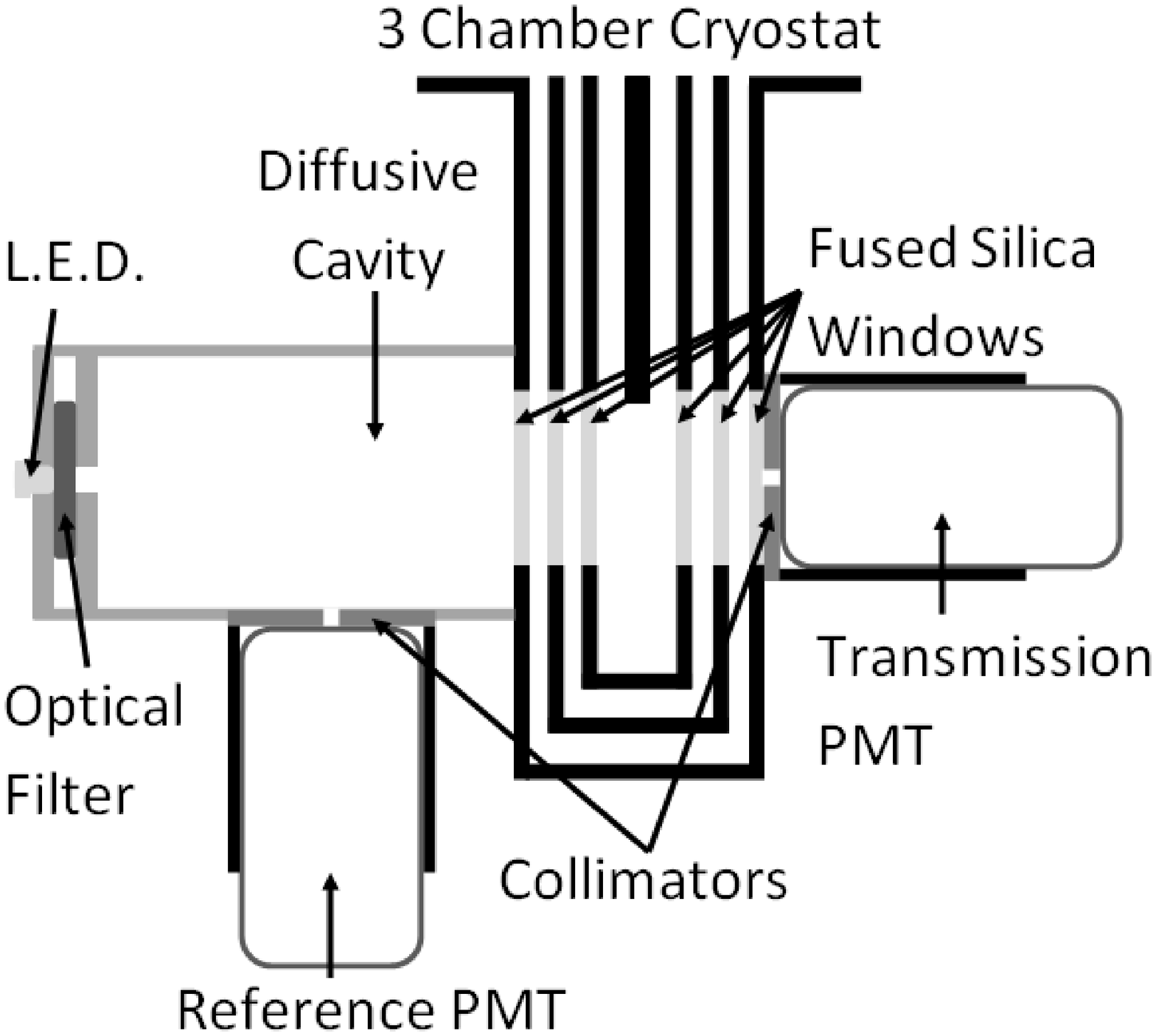,width=\linewidth}
	\caption[Cryostat optical transmission]{Setup to measure the optical transmission of the cryostat (not to scale). At each temperature, a LED, whose  wavelength is selected by an optical filter, shines through the cryostat to the transmission PM.  A  reference PM, fore of the cryostat, monitors the stability of the LED.}
	\label{fig_transmission_setup}
\end{figure}

\begin{figure}[hp]
	\centering
	\epsfig{file=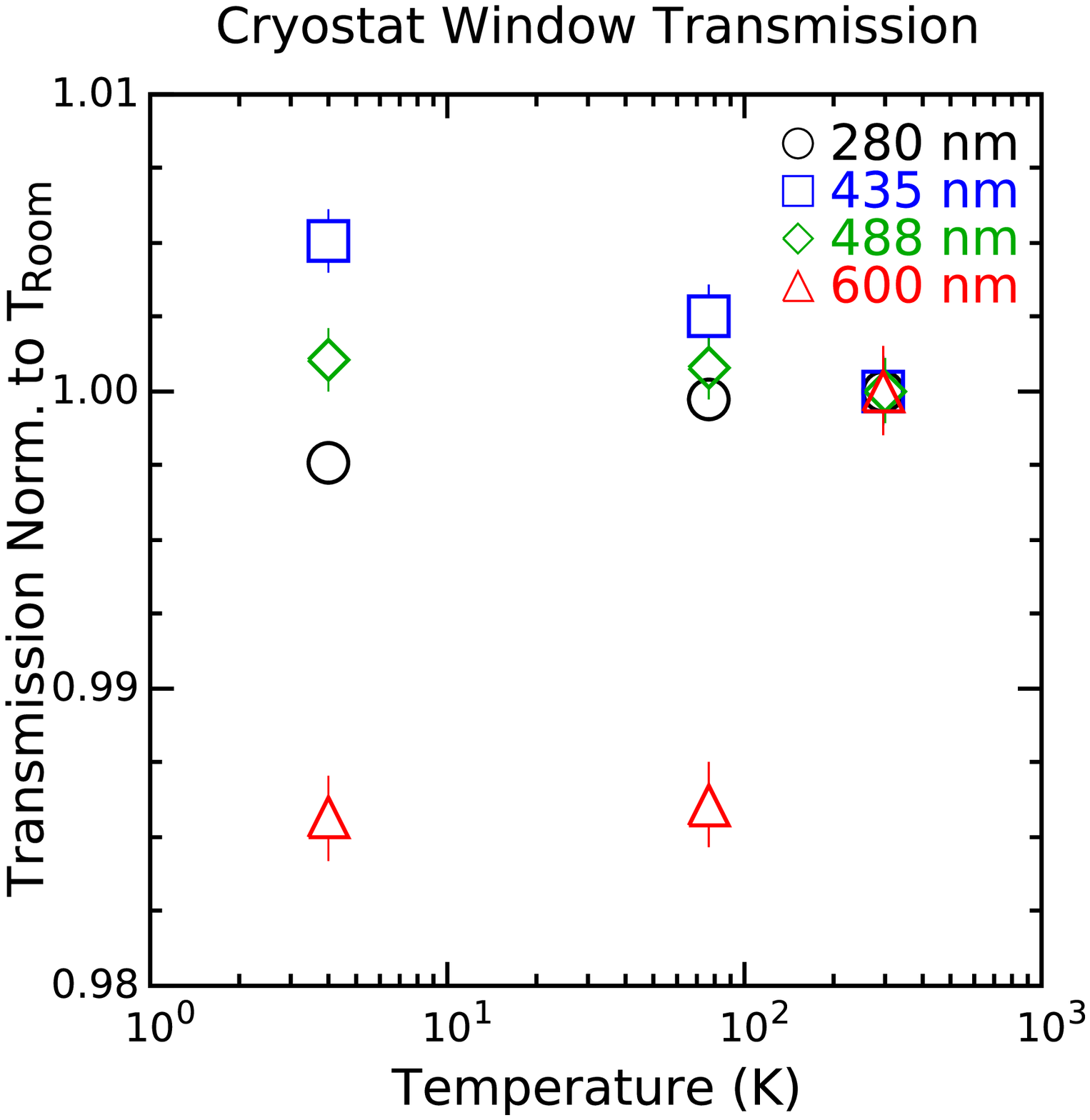,width=1.0\linewidth}
	\caption[Cryostat optical transmission]{Optical transmission of the  cryostat at various temperatures relative to room temperature, for different wavelengths.   At the three shortest wavelengths, transmission varies by less than $0.5\%$ over the temperature range; at the highest wavelength, variation is roughly $1.5\%$. Error bars are one standard deviation statistical ones.  At the highest wavelength, systematic effects may dominate the results.}
	\label{fig_transmission_results}
\end{figure}

We have also used the TDC to study the transmission of the optical cryostat used in these measurements.  The cryostat is closed-cycle, with the compact optical design of a previous one~\cite{verdier_2.8_2009}.  The three sets of windows, at room temperature, $70~\kelv$ and $4~\kelv$ (the last two values are nominal), are made out of fused silica to obtain good transmission over a broad spectral range, and have a nominal transmission of more than $90\%$ between at least $200~\nm$ and $1000~\nm$.  Temperature-induced variations in their transmission could affect the measured light yields at various temperatures.  Therefore, a setup has been built to measure the transmission at a given temperature, relative to the transmission at room temperature.  It is illustrated in Figure~\ref{fig_transmission_setup}.
On one side of the cryostat, along its optical axis, is a LED, followed by an optical filter and a collimator.  Off-axis is a first, collimated,  reference PM that serves to monitor the stability of the LED.  On the other side of the cryostat is an on-axis, collimated photomultiplier that measures the light transmitted through the cryostat.  The collimators and intensity of the LED are chosen so that the PMs see individual photons, at a rate of one to a few $\kHz$.  The PMs are Hamamatsu R7207, which have a broad spectral response (greater than $10\%$ quantum efficiency over the $150~\nm$--$530~\nm$ range), and a low rate of dark counts (nominally below $30~\Hz$).
At each temperature, photons are counted on both PMs with the TDC for a given amount of time, and the ratio of counts is calculated after subtraction of dark counts (the dark counts were measured in a dedicated run with the LED off).  
This provides the relative evolution of the light yield as a function of temperature, but not an absolute measurement of the transmission at each temperature.  Taking the ratio of the transmission and reference PMs cancels out any instabilities in the LED.
Two LEDs have in fact been used, one emitting white light, and one emitting UV light.
To check the stability of the system, a  measurement was taken with each LED and the apparatus on the cryostat at room temperature. 
Over $2.5~\days$ with the white LED,  the ratio of PM counts showed fluctuations with a standard deviation of $0.4\%$ of the mean value. 
This provides an estimate of systematic  instabilities in the system, including effect of precise room temperature on the LED and each PM, and other environmental sources of noise on the PMs.
Measurements have been made at four different wavelengths, using optical filters centered at $280~\nm$ (UV LED), $435~\nm$, $488~\nm$  and at $600~\nm$ (white LED). At each wavelength, measurements were carried out at three temperatures ($289~\kelv$, $77~\kelv$ and $4~\kelv$).  Results are shown in Figure~\ref{fig_transmission_results}. Reported errors come from propagating the statistical error on the number of counts seen on each PM, and are one standard deviation.
For the three lowest wavelength measurements, the transmission varies by less than $0.5\%$ relative to room temperature.  At the longest, orange, wavelength, the variation could be as large as $1.5\%$.  This result in particular may in fact be dominated by systematics mentioned earlier influencing the rates on each PM, since both PMs show a roughly $10\%$ increase in rate at low temperatures, but the increase is slightly greater for the reference PM.
Nonetheless, overall, these variations are small and will have a limited effect on relative light yield measurements of scintillators.

\section{Conclusion}
In the context of a study of scintillators at low temperatures, we have operated a time-to-digital converter (TDC) in streaming mode to identify photons later analyzed offline using the multiple photon counting coincidence  technique to extract timing and light yield information. 
Streams of duration $5$~minutes have been achieved.
Compared to the standard approach that involves a hardware trigger and a digitizer with a fixed acquisition window, being able to chose the coincidence window and acquisition window offline provides greater flexibility and facilitates understanding of spectral features from the scintillator, as demonstrated here with the $60~\keV$ line from \am\ shining on a \zwo\ crystal.
In addition, the amount of TDC data needing to be stored and processed is significantly reduced compared to a digitizer.  
The main drawbacks of the TDC-based method are the slightly greater deadtime between pulses, and the lack of amplitude or integral information --- though neither is generally an issue for the long time constants encountered with many scintillators at low temperatures.  
TDC models with less deadtime between pulses than the model employed here and native ability to stream for hours  exist.
The TDC has also been used to verify that the changes in optical transmission of the cryostat used to cool the samples at various temperatures are small.
The methods described here can be directly extended to double coincidence measurements allowing improved timing accuracy~\cite{verdier_scintillation_2011}, and will be used to further the low-temperature study of scintillators under $\alpha$ and $\gamma$ radiation.

\section{Acknowledgements}
We thank Federica Petricca and Franz Pr\"obst of MPP~Munich for providing the \zwo\ sample.
Sylvie~Chapuy from Agilent~Technologies kindly  commented on the U1051A section of this manuscript.
Summer students Connor~Behan, Ma\"ica~Clavel and Florian~Duquerroix made early contributions respectively to the TDC DAQ, the stream reconstruction software, and the coincidence efficiency studies and transmission measurements.
This work is supported by NSERC Canada (Grant SAPIN No.~386432) and CFI-LOF and ORF-SIF (Project No.~24536).
G.~O. was supported by the Dutch Stichting voor Fundamenteel Onderzoek der Materie (FOM) under Programme~114.

\appendix

\section{Derivation of coincidence efficiencies}
\label{app:CoincProba}
Given $n$ photons arriving on one channel according to an exponential distribution $1/\tau e^{-t/\tau}$, and $m$ photons on the other according to the same distribution (both distributions have the same start time), for a window of given length $T_{coinc}$, what is the probability that it contains at least one photon from each channel (i.e. a coincidence)?

First consider the case of two photons arriving independently at times $t_1$ and $t_2$, one on each channel.  The joint probability density function is:
\begin{equation}
\label{eq:SimpleJointPDF}
\frac{d^2 {\cal P}}{d t_1 t_2} \equiv \frac{1}{\tau^2} e^{-(t_1+t_2)/\tau} H(t_1) H(t_2),
\end{equation}
where $H$ is the Heaviside step function with a value of $0$ for strictly negative arguments and a value of $1$ for positive or null arguments.  Assume that $t_1 < t_2$.  Then the probability that the second photon is not coincident, i.e. that it arrives after time $t_1 +T_{coinc}$, is 
\begin{eqnarray}
\label{eq:SimpleCoincStart}
{\cal P} (t_2 > t_1 + T_{coinc} ) & = & \int_0^{+\infty} \int_{t_1+T_{coinc}}^{+\infty} \frac{d^2 {\cal P}}{d t_1 t_2} d t_1 d t_2 \nonumber \\ & = & \frac{1}{2} e^{-T_{coinc}/\tau}. 
\end{eqnarray}
The same value is obtained assuming that the other photon arrives first ($t_2<t_1$): ${\cal P} (t_1 > t_2 +T_{coinc} ) = \frac{1}{2} e^{-T_{coinc}/\tau}$.  The overall probability of non-coincidence between these two photons is therefore the first probability or the second one, i.e. the sum: 
\begin{eqnarray}
\label{eq:SimpleCoinc11}
p & \equiv & {\cal P} (|t_2 - t_1| > T_{coinc} ) = 2 {\cal P} (t_2 > t_1 +T_{coinc} ) \nonumber \\ & = & e^{-T_{coinc}/\tau}.
\end{eqnarray}

Now assume that one photon arrives on one channel at $t_1$ and $n$ photons arrive on the other at $t_1', t_2', ..., t_n'$.  Assuming all of these events are independent, then there is non-coincidence overall if the $t_1$ is not coincident with $t_1'$ and $t_1$ is not-coincident with $t_2'$ and so forth. In other words, the probability of non-coincidence is the product of the individual probabilities of non-coincidence: $p_{1n} \equiv p^n = e^{-nT_{coinc}/\tau}$.

We lastly consider the case where $n$ photons arrive on the first channel, and $m$ on the second, all independently.  There is non-coincidence overall if the first photon on the first channel is coincident with no photons on the second channel ($p_{1m}$), and the second photon on the first channel is coincident with no photons on the second channel ($p_{1m}$ again), and so forth for all the photons on the first channel.  The overall probability of non-coincidence is therefore $p_{nm} \equiv p_{1m}^n = e^{-nmT_{coinc}/\tau}$.  The probability that there is a at least one coincidence between the two channels is therefore: 
\begin{equation}
\label{eq:SimpleCoincmn}
p_{\mbox{coinc}} = 1-p_{nm} = 1- e^{-nmT_{coinc}/\tau}
\end{equation}
where $\tau$ is the time constant on both channels, $T_{coinc}$ is the coincidence window, and $n$ and $m$ are the number of photons on each channel.

Generalizing this expression to pulse shapes with multiple time constants is straightforward.  For instance, for a probability density function $\sum_{i=1}^{N} \frac{f_i}{\tau_i} e^{-t/\tau_i}$ (where $\sum_{i=1}^N f_i =1$), the joint probability density function becomes:
\begin{equation}
\label{eq:MultiJointPDF}
\frac{d^2 {\cal P}}{d t_1 t_2} \equiv \sum_{i=1}^{N} \frac{f_i}{\tau_i} e^{-t_1/\tau_i} \ \sum_{j=1}^{N} \frac{f_j}{\tau_j} e^{-t_2/\tau_j} H(t_1) H(t_2),
\end{equation}
then, if $t_1 <t_2$, the probability that there is no coincidence is:
\begin{eqnarray}
\label{eq:MultiCoincStart}
{\cal P} (t_2 > t_1 +T_{coinc} ) & = & \int_0^{+\infty} \int_{t_1+T_{coinc}}^{+\infty} \frac{d^2 {\cal P}}{d t_1 t_2} d t_1 d t_2 \nonumber \\ & = & \sum_{i=1}^{N}  \sum_{j=1}^{N} \frac{f_i}{\tau_i} \frac{f_j}{\tau_j} \int_0^{+\infty} e^{-t_1/\tau_i}  \nonumber \\ & \hspace{1cm} & \int_{t_1+T_{coinc}}^{+\infty} e^{-t_2/\tau_j} \ dt_2 dt_1 \nonumber \\ & = & \sum_{i=1}^{N}  \sum_{j=1}^{N} f_i f_j \frac{\tau_j}{\tau_i+\tau_j} e^{-T_{coinc}/\tau_j}. 
\end{eqnarray}
The probability of non-coincidence between two photons, one on each channel, is twice this amount:
\begin{eqnarray}
\label{eq:MultiCoinc11}
p & \equiv & {\cal P} (|t_2 - t_1| > T_{coinc} ) \nonumber \\ & = & 2 {\cal P} (t_2 > t_1 +T_{coinc} ) \nonumber \\ & = & 2 \sum_{i=1}^{N}  \sum_{j=1}^{N} f_i f_j \frac{\tau_j}{\tau_i+\tau_j} e^{-T_{coinc}/\tau_j}.
\end{eqnarray}
By the same reasoning as previously, it follows that the probability of coincidence between $n$ and $m$ photons on each channel is:
\begin{equation}
\label{eq:MultiCoincmn}
p_{\mbox{coinc}} = 1- \left(  2 \sum_{i=1}^{N}  \sum_{j=1}^{N}  f_i f_j \frac{\tau_j}{\tau_i+\tau_j} e^{-T_{coinc}/\tau_j} \right)^{nm}
\end{equation}

\section{Average time and number of photons}
\label{app_avtime_photons}
We first consider a scintillator emitting photons with a single time constant $\tau$. The system detects $n$ of these photons.  The time distribution of these photons will be:
\begin{equation}
\frac{dn}{dt} = H(t) \frac{n}{\tau}e^{-t/\tau}.
\end{equation}
where $H$ is the Heaviside step function.
If $T_{acq} \geq 0$ is the acquisition window starting at $t=0$, then the number of photons actually counted will be 
\begin{equation}
\int_0^{T_{acq}} \frac{dn}{dt} dt = n \left( 1 - e^{-T_{acq}/\tau}\right) \leq n. 
\end{equation}
For an infinite window, this number is $n$.  If the acquisition window is too short, then photons will be missed.  In practice $90\%$ should be counted for $T_{acq}/\tau = 2.3$.

The mean arrival time of photons  is given by: 
\begin{eqnarray}
\frac{ \int_0^{T_{acq}}  t \frac{dn}{dt} dt}{\int_0^{T_{acq}}  \frac{dn}{dt} dt} & = & \frac{n \left( -T_{acq} e^{-T_{acq}/\tau} +\tau \left( 1 - e^{-T_{acq}/\tau}\right)\right)}{n \left( 1 - e^{-T_{acq}/\tau}\right)} \nonumber \\ 
& = & \tau \frac{1 - e^{-T_{acq}/\tau} \left( 1 + T_{acq}/\tau \right)}{ 1 - e^{-T_{acq}/\tau}} \\& \leq & \tau. \nonumber 
\end{eqnarray}
For an infinite window, this yields $\tau$.  For a shorter window, the value will be underestimated.

In the more general case of a scintillator emitting with several time constants, the time distribution of these photons will be: 
\begin{equation}
\frac{dn}{dt} = H(t)\sum_i \frac{n_i}{\tau_i}e^{-t/\tau_i}.
\end{equation}

The number of photons integrated over time $T_{acq}$ will be:
\begin{equation}
\int_0^{T_{acq}} \frac{dn}{dt} dt = \sum_i n_i \left( 1 - e^{-T_{acq}/\tau_i}\right) \leq \sum n_i
\end{equation}
The mean arrival time of photons  is given by: 
\begin{eqnarray}
\frac{  \int_0^{T_{acq}} t \frac{dn}{dt} dt}{\int_0^{T_{acq}}  \frac{dn}{dt} dt} & = &\frac{\sum n_i \left( -T_{acq} e^{-T_{acq}/\tau_i} + \tau_i \left( 1 - e^{-T_{acq}/\tau_i}\right)\right)}{\sum n_i \left( 1 - e^{-T_{acq}/\tau_i}\right)} \nonumber \\ 
& = & \frac{\sum n_i \tau_i \left(  1 - e^{-T_{acq}/\tau_i} \left(1+  T_{acq} / \tau_i \right) \right)}{\sum n_i \left( 1 - e^{-T_{acq}/\tau_i}\right)} \\ 
& \leq & \frac{\sum n_i \tau_i}{\sum n_i} \nonumber
\end{eqnarray}

\end{document}